\documentstyle[psfig]{mn}

\def\kms                 {km\thinspace s$^{-1}$}

\def\etal{{\it et al. }}

\def\Msol{\thinspace\hbox{$\hbox{M}_{\odot}$}}
\def\Zsol{\thinspace\hbox{$\hbox{Z}_{\odot}$}}
\def\sol{\thinspace\hbox{$_{\odot}$ }}

\def\a4{\hsize 17.0cm \vsize 25.cm}

\begin{document}

%

\title[X-rays from superbubbles]
{Evolution of the X-ray luminosity and metallicity of starburst blown 
superbubbles}

%
\author[Sergey Silich et al.]{Sergey A. Silich $^1$
\thanks{RS Visitor at IoA, Cambridge}, Guillermo Tenorio-Tagle $^2$
\thanks{Visitor IoA, Cambridge}, Roberto Terlevich $^3$
\thanks{ Visiting Professor, INAOE}\cr Elena Terlevich $^2$ 
\thanks{Visiting fellow IoA, Cambridge} \& Hagai Netzer $^4$ \\
$^1$  Main Astronomical Observatory National Academy of Sciences of
Ukraine, 03680, Kiev-127, Golosiiv, Ukraine \\
$^2$ Instituto Nacional de Astrof\'\i sica Optica y Electr\'onica, 
AP 51, 72000 Puebla, M\'exico \\
$^3$ Institute of Astronomy, Madingley Road, Cambridge CB3 0HA, UK \\
$^4$ School of Physics and Astronomy, Tel Aviv University, Tel Aviv, Israel.
}


\maketitle

\begin{abstract}
We calculate the time-dependent metal production expected from  starbursts
and use them as  boundary conditions in our 2D simulations of evolving 
superbubbles. We assume  that the produced  metals (oxygen and iron) 
thoroughly mix with the ejected stellar envelopes, and/or with the matter 
thermally evaporated from the superbubble  cold outer shell. The metal 
production process  
determines the time-dependent metallicity in hot superbubble interiors, 
and  leads to values of $Z$ $\geq$ \Zsol when oxygen is used as 
tracer, and  under-solar when the metallicity is measured with respect  
to iron. In either case, the enhanced metallicity 
boosts the  X-ray emissivity of superbubbles, bringing theory and observations 
closer together.  

\end{abstract}


\section{Introduction}

The formation and evolution of galaxies is one of
the most captivating problems of modern astrophysics.  With present-day 
techniques, young star clusters  can be observed  at
large look-back times and much  work has been devoted to the
 detailed studies of the optical and near UV properties of such systems 
 at  intermediate and high redshifts. 
The defining characteristic of starburst galaxies is  their spectrum
which is the emission lines from HII regions in 
 extremely young ($\leq$ 10 Myr) star forming regions.
While the UV radiation from the massive stars excite and ionize the associated 
HII regions, the stellar winds and supernova explosions 
lead to giant hot bubbles in the
interstellar medium (ISM). 

Most of the information about the properties of the ``warm ISM'' in 
young giant star-forming regions comes from the 
analysis of their  emission
lines. Such observations indicate that the
warm ISM in starburst galaxies is not yet contaminated by the metals ejected
by the present burst. 
However, little is known about the ``hot ISM'', where most of the
newly produced metals are presumably located.  
Data from {\it ROSAT\/}, {\it ASCA\/} and {\it BeppoSAX\/},
 although  inconclusive, seem to suggest sub-solar iron
abundance and about solar abundances for the $\alpha$-elements 
(Bauer \& Bregman 1996, Ptak \etal 1999, Persic \etal 1998, but see Dahlem, 
Weaver \& Heckman 1998 and Weaver \etal 1999). However, 
The abundances determined from the optical emission lines, in some of the best
studied 
 galaxies, is considerably higher than those derived from the X-ray data.
This conflict is likely to be resolved soon, with the coming new high 
resolution spectroscopic observations of {\it Chandra} and {\it XMM}.

The mechanical energy from evolving starbursts is  known to 
lead to the formation of superbubbles in the ISM. This happens as the violently 
ejected matter from winds and supernovae becomes thermalized at 
a reverse shock.  This provides it with the high temperature 
($T$ $\geq$ 10$^7$ - 10$^8$ K) and the high thermal pressure that allows it 
to drive a strong shock (the so called outer shock) into the surrounding ISM 
The outer shock is responsible for sweeping  and accelerating the surrounding 
gas into a large-scale shell, while the high sound speed, thermalized ejecta, 
fills most of the volume formerly occupied by the swept up gas. The two gases, 
the ejecta and the swept up ISM, are separated by a contact discontinuity and 
thus there is little enrichment, at this time, of the galaxy ISM 
(Tenorio-Tagle 1996). 

Several 
calculations (see Tenorio-Tagle \& Bodenheimer 1988, Silich \& Tenorio-Tagle
1998 and references therein) 
have shown how the growing structures may acquire a variety of 
shapes, depending on how the ISM is distributed. Fairly round and 
8-shaped remnants are expected for a constant density medium and  plane 
stratified atmospheres, respectively.  Superbubbles are also known to blow out 
upon the sudden acceleration experienced when crossing supersonically a large 
negative density gradient.
 At that moment,
 the accelerated section of the shell
of swept up matter becomes Rayleigh-Taylor unstable and fragments, while 
the hot supperbubble interior is vented into the low density surrounding gas. 
The escaping material, flying with its sound speed, would then push again 
the outer shock to build a new and even larger shell 
evolving into the halos of 
galaxies.  If the shock  reaches the outskirts 
of galaxies with a speed larger than the  escape velocity of the system, 
it may also establish a galactic wind. This is in fact 
detected in the case of some nuclear starbursts (e.g.~Heckman \etal 1996, 
Tenorio-Tagle \& Mu\~noz Tu\~n\'on 1998). The latter event is expected to 
have drastic consequences as  all the  newly processed starburst elements, 
originally found in  the hot superbubble interior,  will  be 
channeled out of their parent galaxy into the 
intergalactic medium. 

Superbubbles  have been recognized for their large-scale 
expanding HI shells (Heiles 1979, Brinks \& Bajaja 1986, Maschenko \etal 
1999). Some of the youngest ones are detected at optical frequencies
 emanating from 
giant HII regions (Meaburn 1980, Heckman \etal 1990, Martin 1996, 
Oey 1996), and many have been recognized by  
the X-ray emitted from their hot interiors (Wang \& Helfand 1991, 
Heckman 1995, Wang 1999). 

Several authors have 
pointed out a large discrepancy between theory and observations of superbubbles. 
Current  predictions  are  based on the fact that,
in the temperature range of $\sim 10^6 - 10^7$ K, the X-ray emissivity
can be  approximated by a linear function of the gas metallicity. For the 
0.1 - 2.4 keV energy band, this  can be approximated by a constant value 
$\Lambda_x = 3 \xi \times 10^{-23}$ erg cm$^3$ s$^{-1}$, where $\xi$
is the metallicity in solar units. 
A simple analytic model may then be developed to estimate the X-ray 
luminosity from a spherically
symmetric, energy dominated bubble bound by a cold radiative shell, and
presenting a 
self-similar temperature and density distributions (see
Chu \& Mac Low 1990 for a constant energy input rate, and Silich 1995 for
a power-law energy deposition). For a constant ambient gas density,
and constant energy input rate, the  X-ray luminosity 
over the energy band 0.1 - 2.4 keV 
 is given by 
\begin{equation}
      \label{eq.0}
L_x = 10^{36} \xi I(\tau) L_{38}^{33/35} n_o^{17/35} t_7^{19/35} erg \,
      s^{-1},
\end{equation}
where L$_{38}$ is the  mechanical luminosity of the starburst
in units of  10$^{38}$ erg s$^{-1}$, 
n$_0$ is the ambient gas number density, and t$_7$ is the evolutionary time in
units of 10$^7$ yr. I($\tau$) is a dimensionless integral:
\begin{equation}
      \label{eq.01}
I(\tau) = \frac{125}{33} -5 \tau^{1/2} + \frac{5}{3} \tau^3 -
          \frac{5}{11} \tau^{11/2}, 
\end{equation}
and $\tau = T_{cut}/T_c$ is the ratio of the X-ray cut-off temperature
(or lowest limit considered for the gas temperature to lead to an important 
X-ray emission; $T_{cut} \approx 5 \times 10^5$ K) to the bubble central 
temperature.

These simple considerations have been applied to a number 
of objects, from
stellar wind bubbles to starburst galaxies. In many cases, the observed
X-ray luminosities are in clear disagreement with the model
predictions and exceed, by much, the model predictions
 (e.g. Chu \& Mac Low, 1990 for LMC bubbles; Walter et al., 1998 for a
superbubble in the nearby dwarf galaxy IC 2574: Martin \& Kennicutt,
1995 for diffuse X-ray emission from the central region of the
starburst galaxy NGC 5253). 
There are several proposed solutions.
 Chu \& Mac Low (1990) have proposed off-centered
supernova explosions. Franco et al. (1993) included the interaction of a 
fragmented ejecta with the outer shell and Martin \& Kennicutt (1995) concluded
that cloud evaporation may be a dominant mechanism to increase the X-ray
emission. 

Here we take the model one step further and show
that the injection of energy and mass from an aging stellar 
cluster not only leads to the large-scale evolving superbubbles 
detected at a variety of frequencies, but also to a time-dependent enhanced 
metallicity  of their hot interior.  Section 2 discusses the time dependent 
production of metals released within the hot superbubble interior. Section 3 
presents the calculations of the  energy deposited by massive coeval 
starbursts and the implications regarding the cooling of the gas and the 
mixing of the
enriched material. The resultant time-dependent X-ray luminosity, is given in 
section 4 and some conclusions are drawn in section 5. 

\section{Time-dependent metal production in starbursts}
 
We assume that the total amount of gas ejected by supernovae (SNe) includes an 
important fraction of newly synthesized oxygen and iron, and that the 
extended stellar outer envelope has the same metallicity as the host galaxy 
ISM. We further assume that the  matter 
violently ejected by SN as a starburst evolves is efficiently thermalized at a 
reverse shock and well mixed with the gas that evaporates from the cold 
radiative 
shell segments. However, a contact discontinuity 
inhibits its immediate mixing with the ISM, which is 
rapidly removed and accelerated by the outer shock (Tenorio-Tagle 1996). 
Consequently, the metals ejected by sequential SNe are to be found in the 
``hot cavity" or  superbubble interior, causing drastic changes to its 
metallicity.  
 
We assume a Salpeter stellar initial mass distribution
(IMF)   
\begin{equation}
      \label{eq.1}
n(m) = f_0 m^{-\alpha},
\end{equation}
within a range of upper $M_{up} = 100 M_{\odot}$ and  lower 
$M_{low} = 1M_{\odot}$ cut-off masses and a slope of $\alpha$ = 2.35. The 
normalization constant $f_0$
is determined by the total mass of the star cluster $M_{SB}$,
\begin{equation}
      \label{eq.2}
f_0 = \frac{(\alpha-2) M_{SB}}{M_{low}^{2-\alpha} - M_{up}^{2-\alpha}}.
\end{equation}
Assuming for simplicity that massive stars loose {\it all their mass} as they 
explode as SNe
(Pilyugin 1992, Pilyugin \& Edmunds 1996),  we find the 
 total  ejected mass as a function of the cluster age $t$,
\begin{equation}
      \label{eq.3}
M_{ej}(t) = f_0 \int_{M_{\star}(t)}^{M_{up}} m^{1-\alpha} {\rm d}m =
M_{SB} \frac{M_{\star}(t)^{2-\alpha} - M_{up}^{2-\alpha}}
                     {M_{low}^{2-\alpha} - M_{up}^{2-\alpha}},
\end{equation}
where the mass of the stars exploding after an evolutionary  time $t$,  ($M_{\star}(t)$), 
has been 
found using Chiosi et al. (1978) and Stothers (1972) approximations to the 
main sequence lifetime of massive stars:
\begin{equation}
      \label{eq.4}
M_{\star}(t)  = \left\{
\begin{array}{lcl}
10(9 \times 10^6 / t)^2 M_{\odot}, \hfill \ \ {\rm for} \ \  
                 30M_{\odot} \le M_{\star} \le 100M_{\odot}
\\ [0.2cm]
10(3 \times 10^7 / t)^{5/8} M_{\odot}, \hfill \ \ {\rm for} \ \  
                 7M_{\odot} \le M_{\star} \le 30M_{\odot}.
\end{array}
\right.
\end{equation}

Heavy element yields from massive stars have been considered in a number 
of papers (e.g. Renzini et al. 1993).  
However,  the exact values 
 depend strongly on the adopted stellar evolution models. A  recent 
attempt to incorporate heavy elements injection into hydrodynamical bubble 
models, by D'Ercole \& Brighenti (1999), has included only averaged values. 
We have decided to consider several possible scenarios. 
The  oxygen yield as a function of the stellar-mass ($M_{\star}$) 
can be approximated by  two 
different tracks: one is the ``no wind'' (NW) Pilyugin \& Edmunds (1996) analytic
 approximation to Maeder (1992) and Thielemann \etal  (1993) 
models for the yield

\begin{equation}
      \label{eq.5}
Y_O(M) = 0.094(M-10.5)^{1.272} \, M_{\odot},
\end{equation}
which  neglects the oxygen yield for  stars with a mass smaller than 10.5 M$_{\odot}$. The 
second 
 ``with wind" (WW) approximation,  follows from stellar evolution models 
accounting for stellar winds (Maeder 1992, Woosley \etal 1993). In this case 
equation (\ref{eq.5}) can be used within the 10.5M$_{\odot} \le M_{\star} \le 
25M_{\odot}$ 
range, assuming a constant yield up to the upper mass limit  
M$_{up} = 100$M$_{\odot}$.

The long term iron contamination comes mainly from the SNIa, which produce 
0.5--0.7 M$_{\odot}$ of iron after the $^{56}$Ni decay. However, we are 
interested here in the earlier stages of the starburst evolution and thus the 
iron yield from the SNIa has not been taken into account. The 
iron production from type II SN is highly uncertain (see Renzini \etal 
1993) and,  therefore, we have considered two extreme models: Arnett (1991) and 
Thieleman \etal (1992). 
In the Arnett (1991) model, the iron yield increases  with the mass of the 
star, and can be approximated by a linear function  for stars between  
 10M$_{\odot} \le M_{\star} \le 40M_{\odot}$.

\begin{equation}
      \label{eq.6}
Y_{Fe}(M_{\star}) = 0.02 + 0.006(M_{\star}-10) \, M_{\odot},
\end{equation}
For other stellar masses we have assumed constant values of $Y_{Fe}$.  
$Y_{Fe} =$ 0.02 M$_{\odot}$, for $M_{\star} \le 10M_{\odot}$. 
and $Y_{Fe} =$ 0.2 M$_{\odot}$ for $M_{\star} > 40M_{\odot}$.
In Thielemann et al. (1992), the iron yield is 
approximated by the exponential function 
\begin{equation}
      \label{eq.7}
Y_{Fe}(M_{\star}) = \frac{0.423}{\exp{[0.31(M_{\star}-10.5)}]} + 0.045 \, 
M_{\odot},
\end{equation}
within the 13M$_{\odot}$ - 25M$_{\odot}$ range, and assumed to be constant 
outside this range and equal to   $Y_{Fe}(M_{\star}) = 0.24$ M$_{\odot}$ for 
low mass stars, and  $Y_{Fe}(M_{\star}) = 0.05$ M$_{\odot}$ for stars with an 
initial mass larger than 25 M$_{\odot}$. We shall refer to the above 
approaches for deriving the iron production rates as A (for 
Arnett 1991), and T (for Thielemann \etal 1992), respectively. 

We have also assumed that the stellar matter ejected 
 before the first SN explosion ($\sim 3$ Myr)  
 has the same metallicity as the ISM 
(see table 1).  Note that the expressions used for the yields are for 
solar metallicity stars (which are the ones available) while the stars in the 
models have metallicities of 0.1 \Zsol . In what follows, the integrated metal 
content of superbubbles is given in solar values for which Z$_O$ = 0.0083 and 
Z$_{Fe}$ = 0.00126 (Grevesse, Noels \& Sauval 1996).  

The metallicities of the hot bubble interior relative to  solar are 
given by
\begin{eqnarray}
      \label{a.1}
      & & \hspace{-0.5cm}
Z_{hot,O} = \frac{M_{ej,O}/Z_{O} + Z_{ISM} M_{ev}}
                        {M_{ev} + M_{ej}},
      \\[0.2cm]
      & & \hspace{-0.5cm}
Z_{hot,Fe} = \frac{M_{ej,Fe}/Z_{Fe} + Z_{ISM} M_{ev}}
                        {M_{ev} + M_{ej}},
\end{eqnarray}
where M$_{ev}$ is the mass added to the shocked wind region due to the
cold outer shell evaporation.
  
\begin{figure}
\psfig{figure=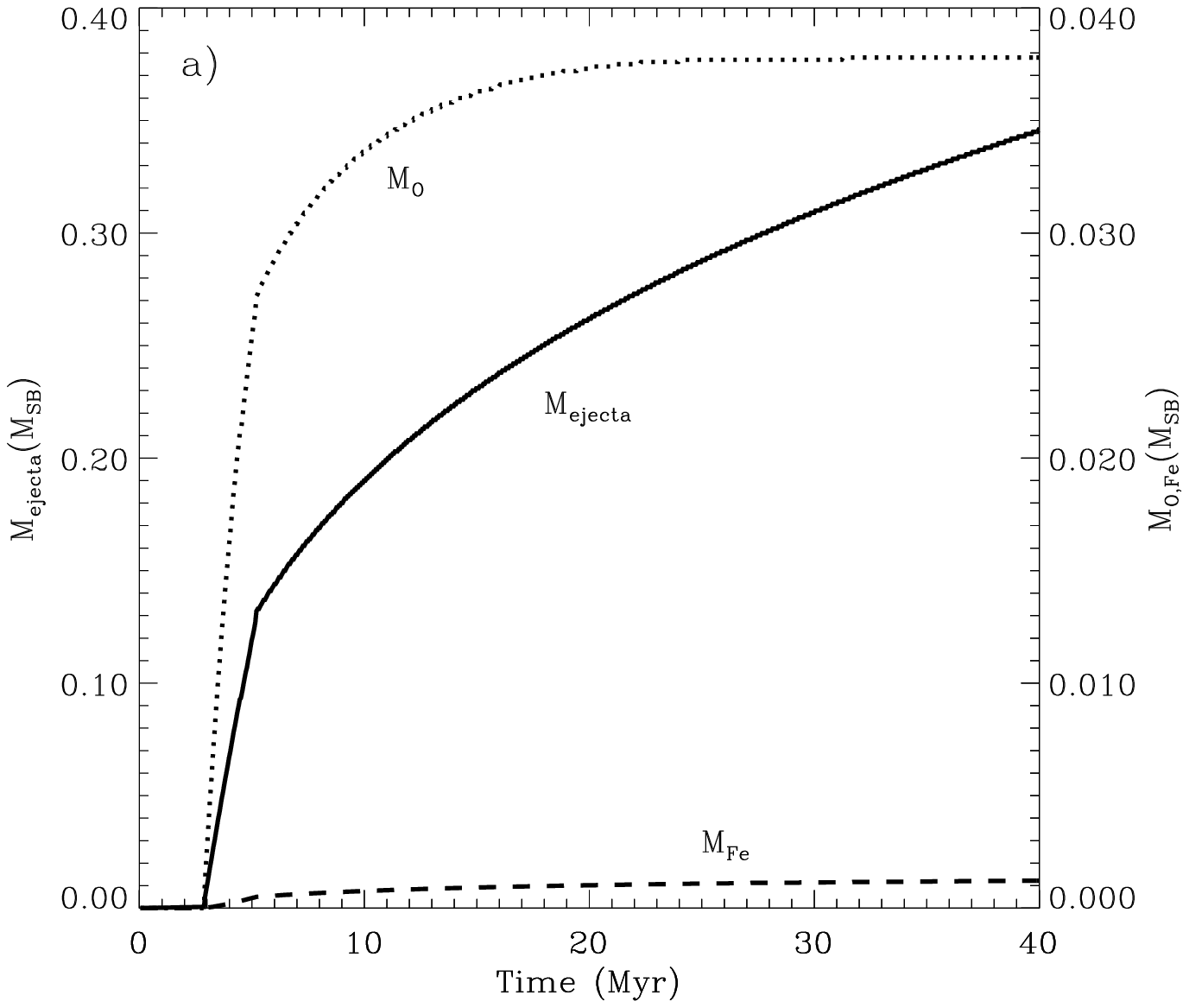,height=8cm,width=8cm,angle=0}
\psfig{figure=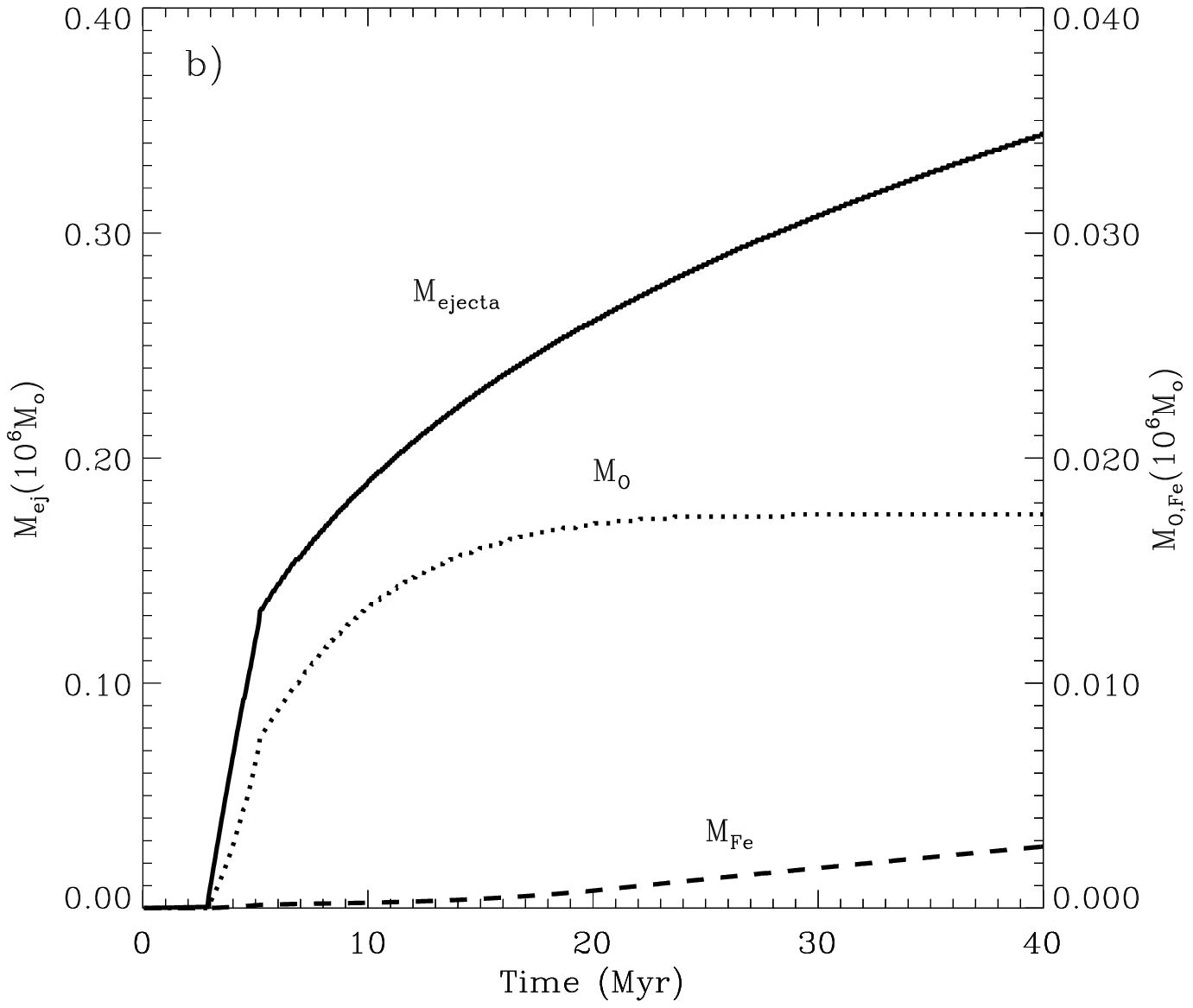,height=8cm,width=8cm,angle=0}
\caption
{Time-dependent metal production from starbursts.
The panels display the cumulative amount of matter (in units of the starburst 
mass $M_{SB}$) ejected by SNe
as a coeval starburst evolves in time (solid line). Also shown are the
amounts corresponding to oxygen (dotted lines) and to iron (dashed lines). 
{\bf (a)}
The oxygen production values were derived under the NW assumption  and
those for iron following the A approximation. {\bf (b)}
The corresponding values  assumed the WW and the T approximations for oxygen 
and iron.
respectively. The right hand side ordinate indicates the mass of O or Fe in 
units of the starburst mass}
\label{fig1} 
\end{figure}

Figure~\ref{fig1} shows the time dependent total 
amount of matter injected into the superbubble 
interior 
($M_{ejecta}$) as a consequence of type II supernova for the NW and 
A approximations and the WW and T assumptions (panels {\bf a} and {\bf b}, 
respectively). This is to be compared with the total amount of oxygen 
($M_O$) and iron ($M_{Fe}$) produced by the star cluster during the first
40 Myr of evolution. Note that at this time the amount of iron has not yet 
reached its final value, as an important contribution is expected later on 
from stars with a mass smaller than 10 M\sol.   

\section{Superbubbles evolution}

We have carried out calculations in order to find  the impact 
that the metals ejected by a 10$^6$ M\sol coeval starburst may have
 on the ISM of galaxies with low metallicity 
($Z$ $\sim$ 0.1 Z\sol). The calculations were carried out with our three 
dimensional Lagrangian code, based on the thin layer approximation 
(Bisnovatyi-Kogan \& Silich 1995, Silich \& Tenorio-Tagle 1998). 
In all cases the total mass of the host galaxy amounts to 
10$^{10}$  \Msol, while the gas mass amounts only to 10$^9$ M\sol. The galaxy 
model was approximated with a similar prescription to that used by Li \& 
Ikeuchi (1992), Tomisaka \& Bregman (1993), Silich \& Tenorio-Tagle (1998) 
and Tenorio-Tagle \etal (1999; hereafter paper I). The gas 
density distribution allows for two isothermal components. One  is 
related to a central dense molecular disk with a 
mass of 5 $\times $ 10$^7$ M\sol and the second represents the low 
density, extended neutral halo. Both components are in a quasi-equilibrium 
state supported by rotation and random gas motions with a velocity dispersion 
of 20 and 80 \kms, respectively. The temperature of both 
components is  $\sim 1000$ K.

\begin{figure}
\hskip 3.5cm\psfig{figure=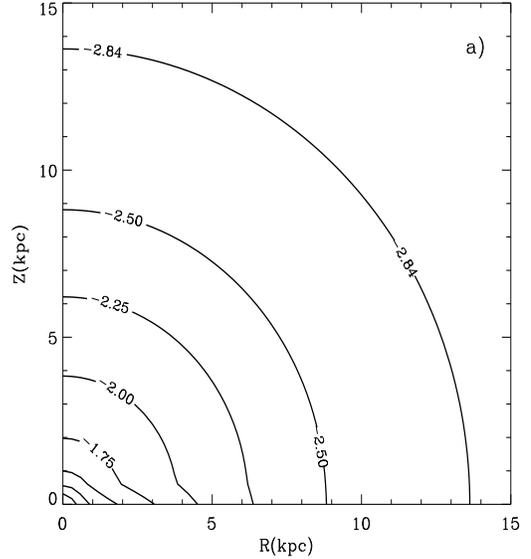,height=8cm,width=10cm,angle=0}
\caption
{The logarithmic density distribution for a 10$^9$ M\sol ISM model. Labels 
correspond to the densities and the outer contour represents the edge of the 
galaxy.}
\label{fig2} 
\end{figure}

 The assumed
 initial density distribution of the model (Figure~\ref{fig2}) is similar to 
that of 
case A1 in Paper I. Despite its large column density ($\sim$ 10$^{23}$ 
cm$^{-2}$) only a small fraction of this mass arises from the extended halo, 
and  most of the undisturbed column density is  within the dense 
molecular core. The 10$^6$ M\sol starburst was assumed to produce a constant 
mechanical energy at a rate of $\sim$ 3.2 $\times$ 10$^{40}$ erg s$^{-1}$ (see 
Leitherer \& Heckman 1995 synthesis models for low metallicity starbursts) for 
the first 40 Myr of its evolution, leading to a giant evolving superbubble.  
Three different cases are presented. In all we assume that the 
supernova products (oxygen and iron) rapidly mix with all the gas 
 within the superbubble interior. As discussed in section 2, two 
different metal production rates, for both oxygen and iron were used, 
and their possible impact is discussed here and in section 4.

Case 1 follows  the amount of mass thermally evaporated from 
the cold shell segments and  considers its  mixing with the ejecta 
from type II supernovae. The
NW and A  approximations were used for case 1A, and the WW  and
T approximations, for case 1B. 
Case 2 neglects the process of thermal evaporation and
thus the metals are allowed to mix only with the ejected stellar envelopes,
the latter assumed to have the same metallicity as the ISM in the host galaxy. This 
calculation leads  to the largest metal rich superbubbles.
Case 3 is almost identical to case 1, with the exception that the assumed  
mechanical energy input rate is the one expected from a 10$^7$ M\sol 
coeval starburst (3.2 $\times$ 10$^{41}$ erg s$^{-1}$) instead of the previous 
10$^6$ M\sol ones. Case 3A follows the bubble evolution within a low 
metallicity ISM (Z=0.1\Zsol ), whereas  case 3B assumes solar metallicity 
for the host galaxy ISM. The model input parameters are summarized 
in Table~\ref{models}.

\begin{table*}
\centering
\begin{minipage}{140mm}
  \caption[model parameters]{Model parameters}
\begin{tabular}{@{}lccccc@{}}
Model & L$_{SB}$ & Y$_O$ & Y$_{Fe}$ & Z$_{ISM}$ & Shell evap. \\
& erg s$^{-1}$   &       &          & Z$_{\odot}$ & \\[10pt]
1A & 3.2 $\times 10^{40}$ & NW & A & 0.1  & yes  \\
1B & 3.2 $\times 10^{40}$ & WW & T & 0.1  & yes  \\
2A & 3.2 $\times 10^{40}$ & NW & A & 0.1  & no  \\
2B & 3.2 $\times 10^{40}$ & WW & T & 0.1  & no  \\
3A & 3.2 $\times 10^{41}$ & NW & A & 0.1  & yes  \\
3B & 3.2 $\times 10^{41}$ & NW & A & 1.0  & yes  \\
A1 & 3.2 $\times 10^{40}$ & - & -  & 0.1  & yes  
\end{tabular}
\label{models}
\end{minipage}
\end{table*}


The calculations follow the change of the metallicity in the superbubble interior
with time  
 and apply the correspondingly modified  cooling rates. 
For the outer shell we assumed that its metallicity remains  constant in time 
and equal to Z$_{ISM}$. The results are compared to our case  A1  from paper 
I, which assumed  that the cooling function, scaled to the initial metallicity 
of the host galaxy, remained unchanged with time, despite the obvious injection
of new metals into the superbubble interior. Furthermore, our comparison 
case A1 also assumed, as in most calculations in the literature, 
that the same cooling function could be applied throughout the flow.

\begin{figure}
\psfig{figure=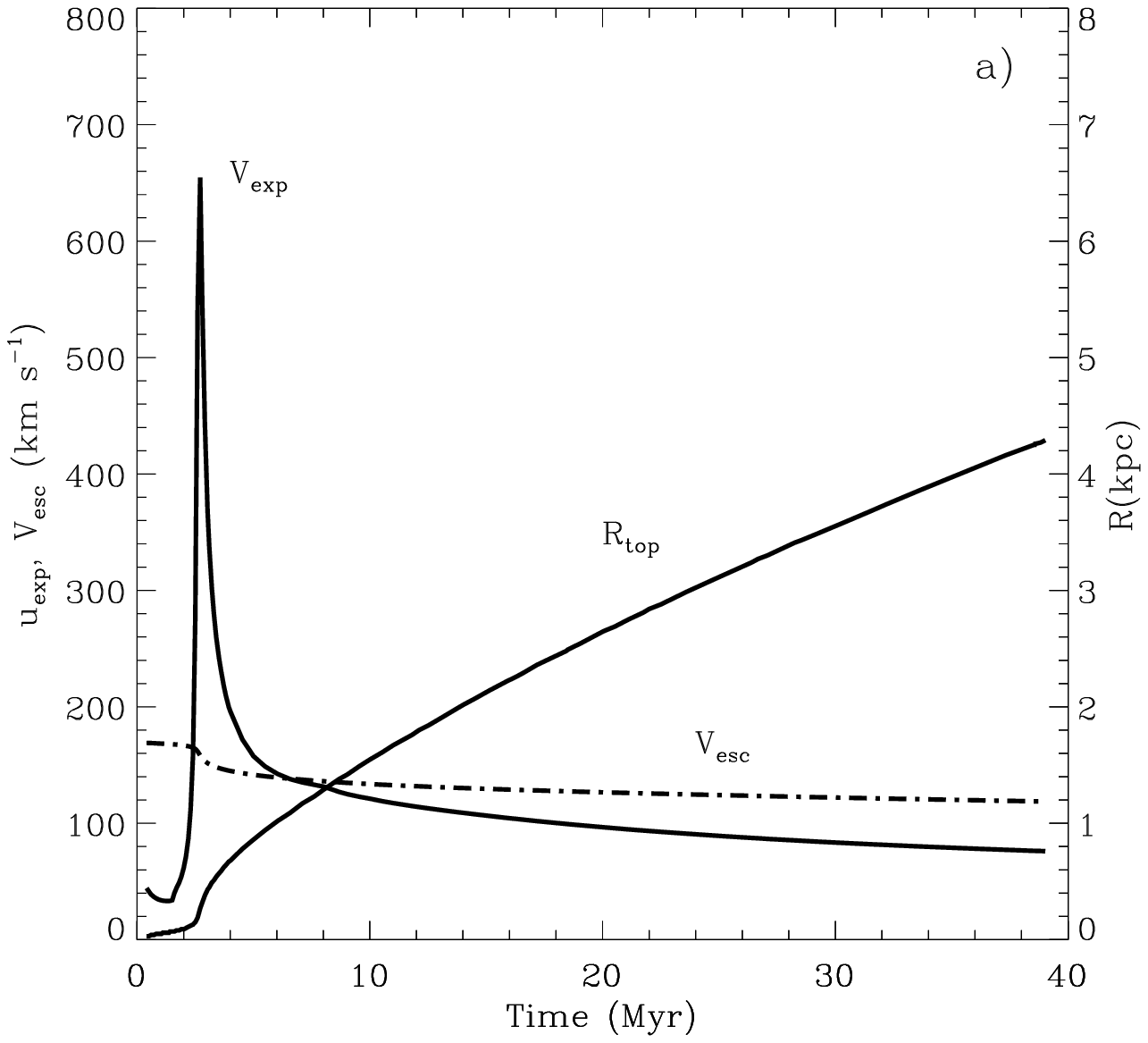,height=8cm,width=8cm,angle=0}
\psfig{figure=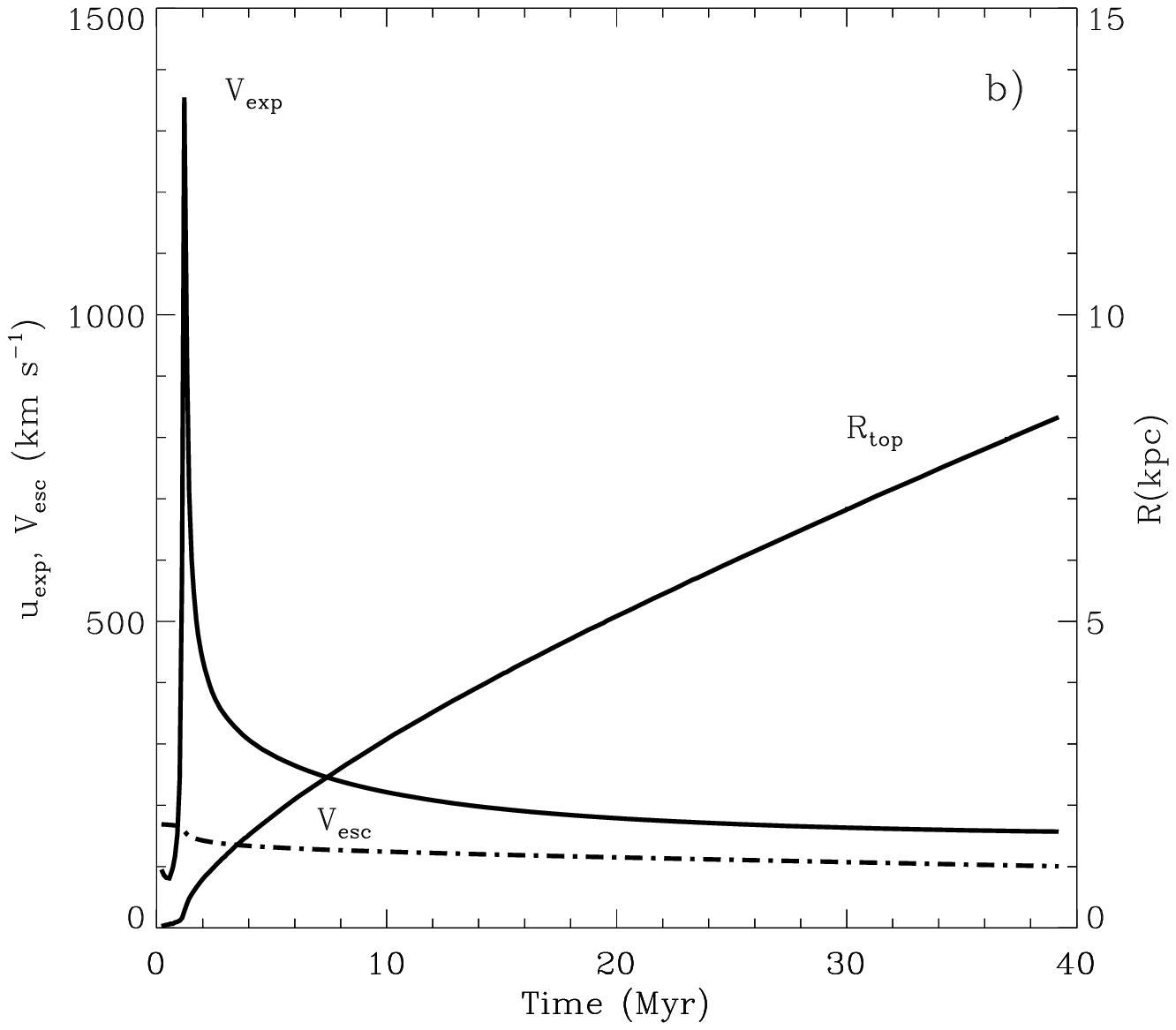,height=8cm,width=8cm,angle=0}
\caption
{Evolution of the superbubbles. {\bf (a)} Case 1A  
and {\bf (b)} case 3A; shown as a function of time are the
 maximum expansion speed (\kms) measured along the symmetry axis 
(solid lines), the  galaxy escape velocity (dash-dotted lines) and the radius 
(in kpc) reached by the superbubble.}
\label{fig3} 
\end{figure}

Figures~\ref{fig3}a and~\ref{fig3}b, show the evolution with time of the 
size and expansion velocity of the
superbubble for cases 1A and 3A,
respectively. The figure  displays values of the fastest expansion velocity 
measured along the symmetry axis, as well as of the largest radius 
acquired by the superbubble during the first 40~Myr of evolution. The maximum 
expansion velocity shows  an initial deceleration, followed by a 
strong acceleration that leads to several hundreds of \kms\ immediately after
blow-out from the central disk. This short phase is abruptly 
interrupted once sufficient halo matter has been swept into the expanding 
shell causing, once again, a steady deceleration of the remnants.  
The maximum expansion speed is to be compared with the escape velocity from 
the host galaxy  (V$_{esc}$), to discern whether  the 
decelerating remnant remains bound (as in case 1) or is to reach the 
galaxy outer boundary and eject its contents into the intra-cluster medium (as 
in the more energetic case 3).  The results of cases 1, 2 and those of case A1 
of paper I are identical, as all three  assumed the same mechanical
energy injection and the same galaxy. The generated remnants, 
in all three, have  
identical size, expansion speed and  total amount of swept-up mass.  
Below we show that,  on the other hand,  the total luminosity and the metal 
content inside 
the superbubble, are different for the different models.

\begin{figure}
\psfig{figure=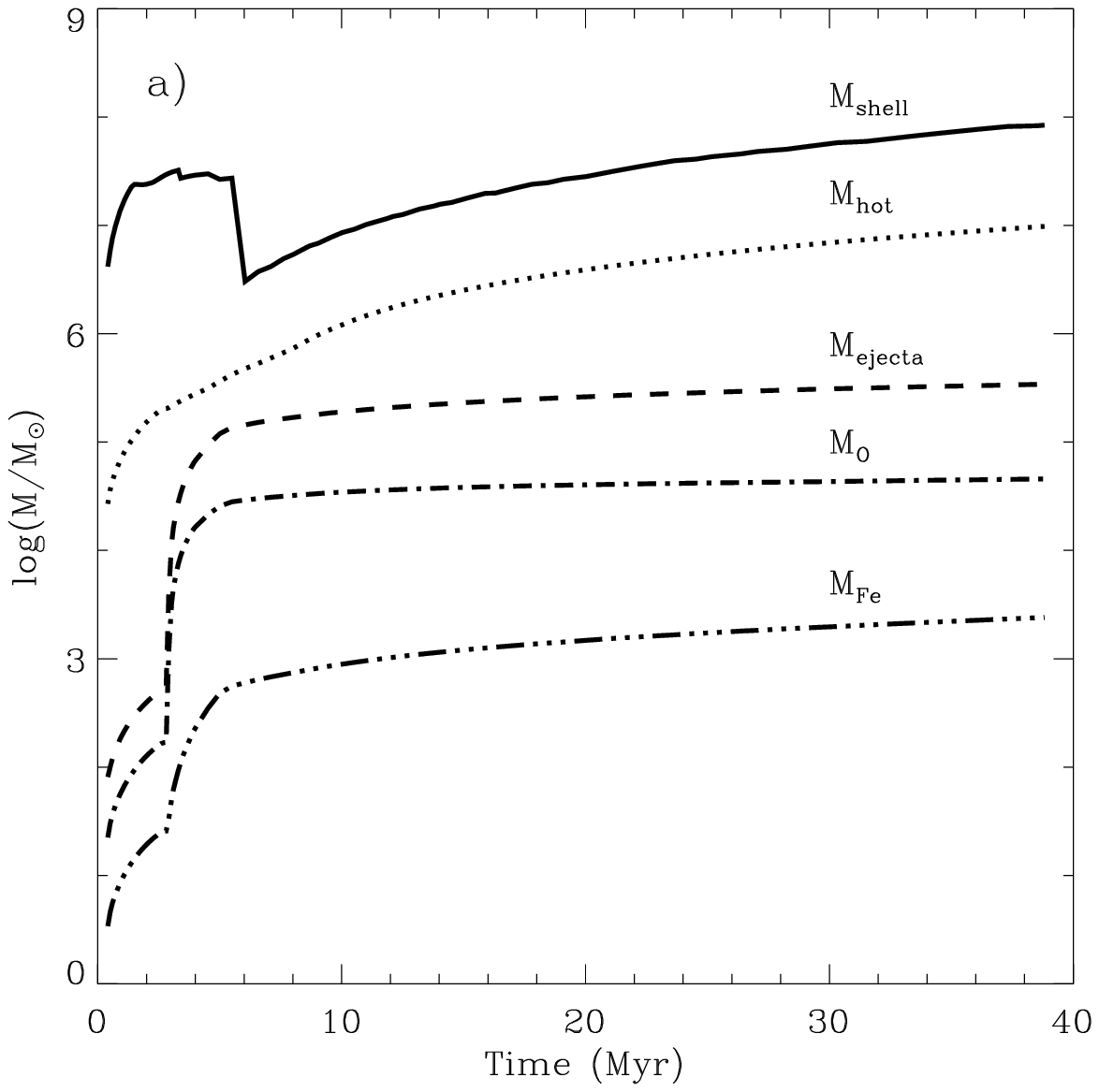,height=8cm,width=8cm,angle=0}
\psfig{figure=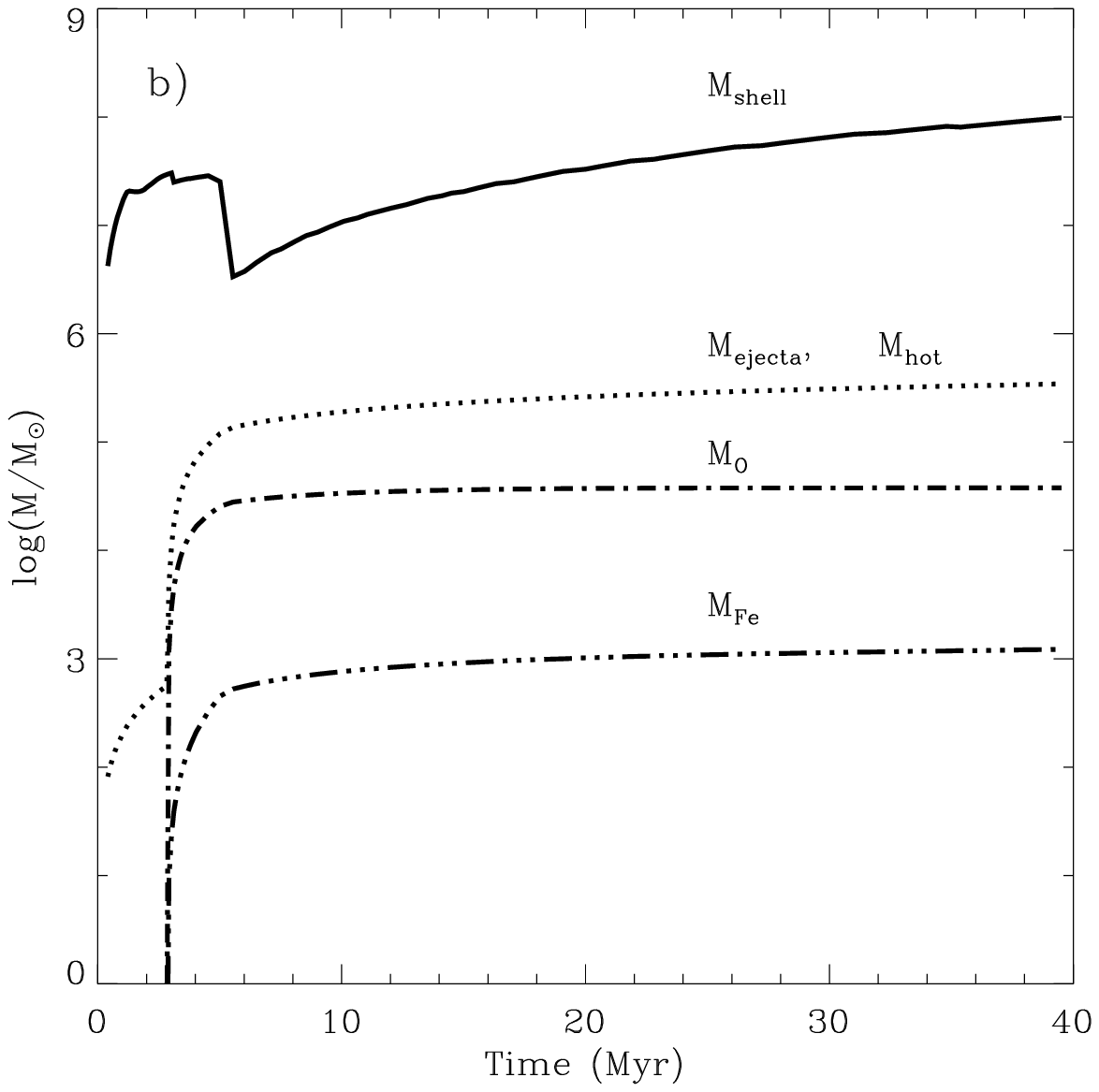,height=8cm,width=8cm,angle=0}
\caption
{The log of the swept up mass (in \Msol) and that inside the
superbubbles for cases 1A {\bf (a)} and 2A {\bf (b)}, as a function of time. 
The curves represent the log of: total swept up mass (solid lines), total 
mass inside the superbubbles (dotted lines labeled $M_{hot}$). The 
latter includes both the amount of matter thermally evaporated from the shell
(not shown) and the mass ejected by SN (dashed lines labeled $M_{ejecta}$, coinciding with $M_{hot}$ in panel b).
Also shown are the log of the total oxygen and iron mass (dash-dotted lines).  
In both cases these were calculated following the NW and A
approximations (see  section 2)}
\label{fig4} 
\end{figure}

The results for cases 1A and 2A are shown in  Figure~\ref{fig4}. The diagram 
shows the 
amount of matter swept by the outer shock ($M_{shell}$), 
and the mass of thermally evaporated gas from the outer shell  
that is ejected by supernovae ($M_{hot}$) 
(all of which is to be found in the hot superbubble 
interior) as a function of time. Also shown are the 
cumulative amounts of 
matter ejected by SNe ($M_{ejecta}$) and the corresponding  amounts 
of  oxygen  ($M_{O}$) and iron ($M_{Fe}$). 
The sharp  $M_{shell}$  decrease observed at about 5~Myr results from the 
outer shock merging at the mid-plane of the galaxy (see  Silich \& 
Tenorio-Tagle 1998). 

In case 1 (Figure~\ref{fig4}a), the injected metals were allowed to  mix 
thoroughly
with the matter evaporated from the expanding shell. Note that for such a 
massive starburst, the amount of hot matter evaporated from the shell reaches 
a large proportion of the total mass of the shell  
($M_{hot} \sim$ 10$\%$ $M_{shell}$). The total oxygen mass surpasses the  
10$^4$ M\sol value while the total mass in iron is $\sim$ 10$^3$ M\sol. 

Case 2 (Figure~\ref{fig4}b) assumes no mass evaporation from the 
shell, and thus apart 
from the small contribution due to the early wind phase, the amount of 
$M_{hot}$ is very similar to the  matter ejected by supernovae ($M_{ejecta}$). 
Thus the amount of oxygen and iron are  very close
to the values  expected from the synthesis caused by the massive starburst. 

\begin{figure*}
\centerline{
\hbox{
\psfig{figure=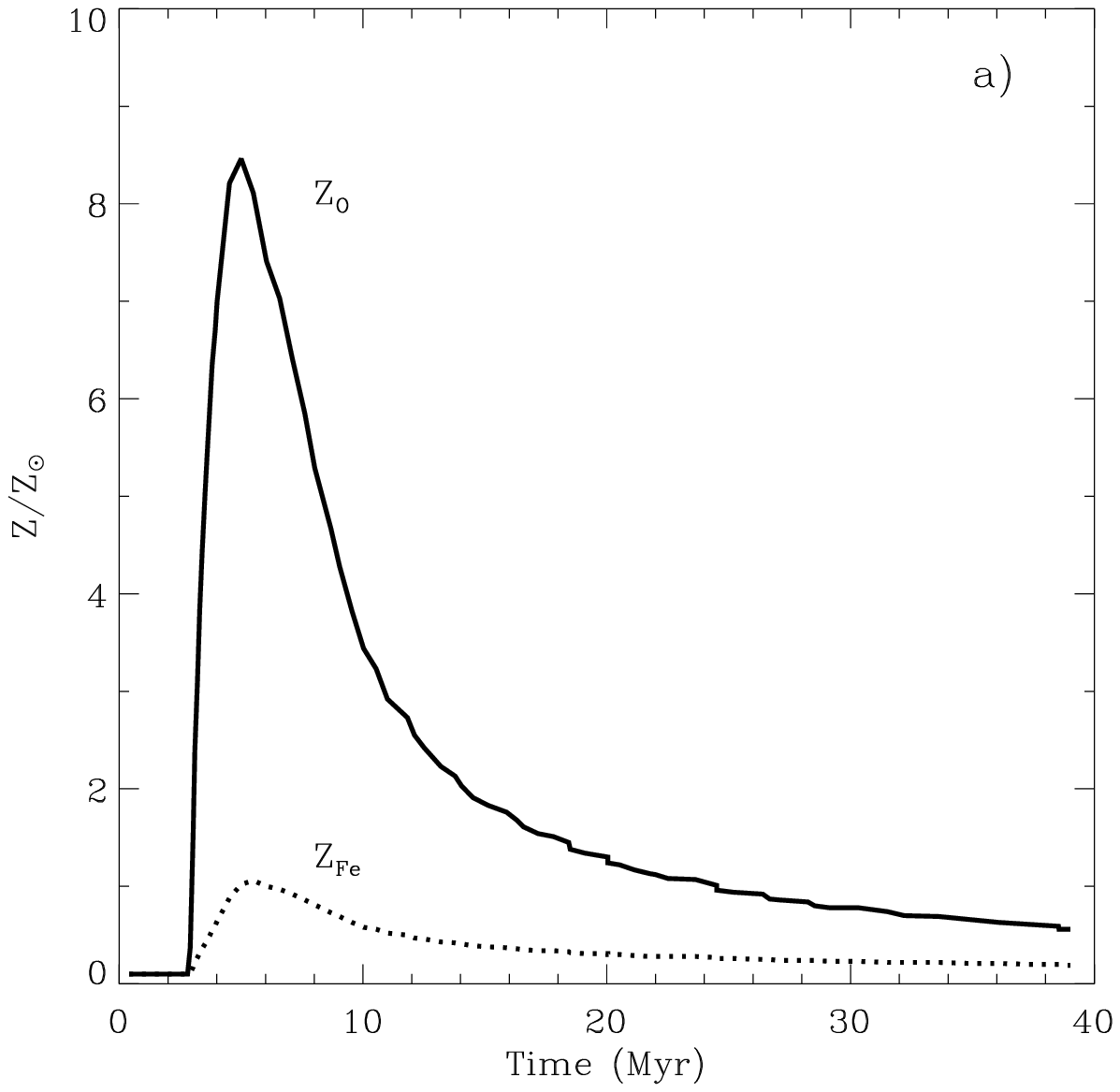,height=7cm,width=8cm,angle=0}
\hspace{2mm}
\psfig{figure=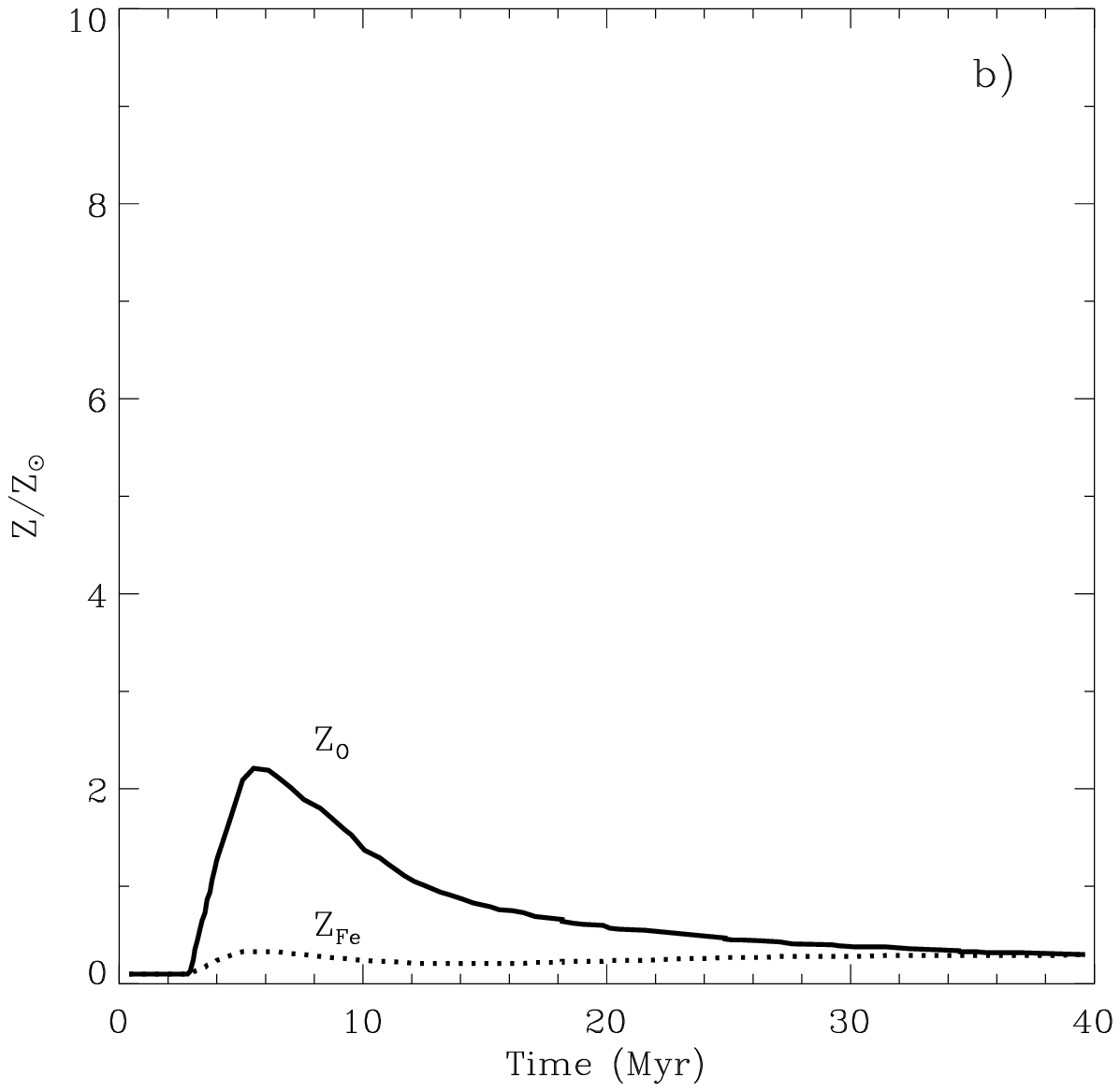,height=7cm,width=8cm,angle=0}
}}
\centerline{
\hbox{
\psfig{figure=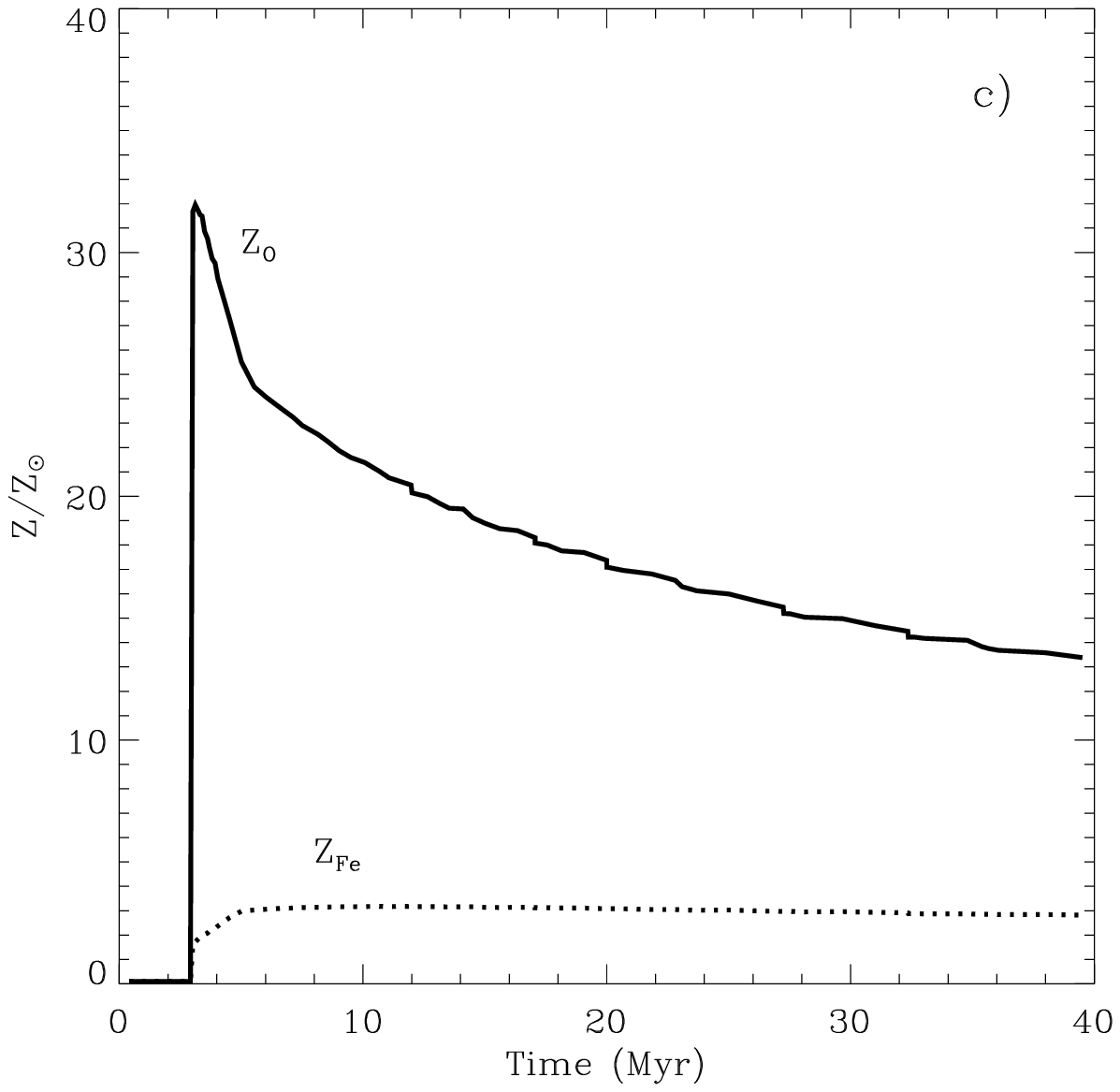,height=7cm,width=8cm,angle=0}
\hspace{2mm}
\psfig{figure=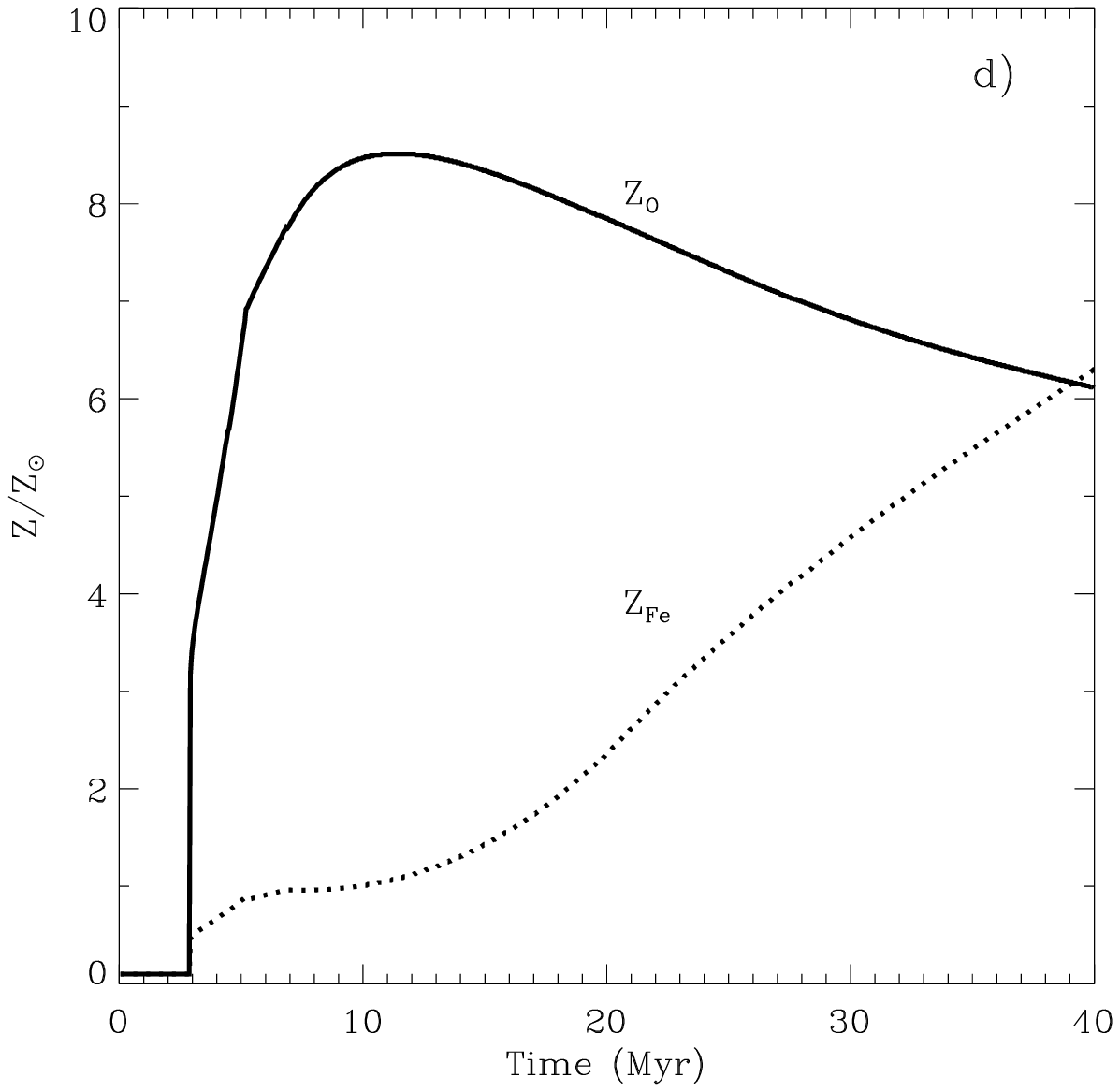,height=7cm,width=8cm,angle=0}
}}
\centerline{
\hbox{
\psfig{figure=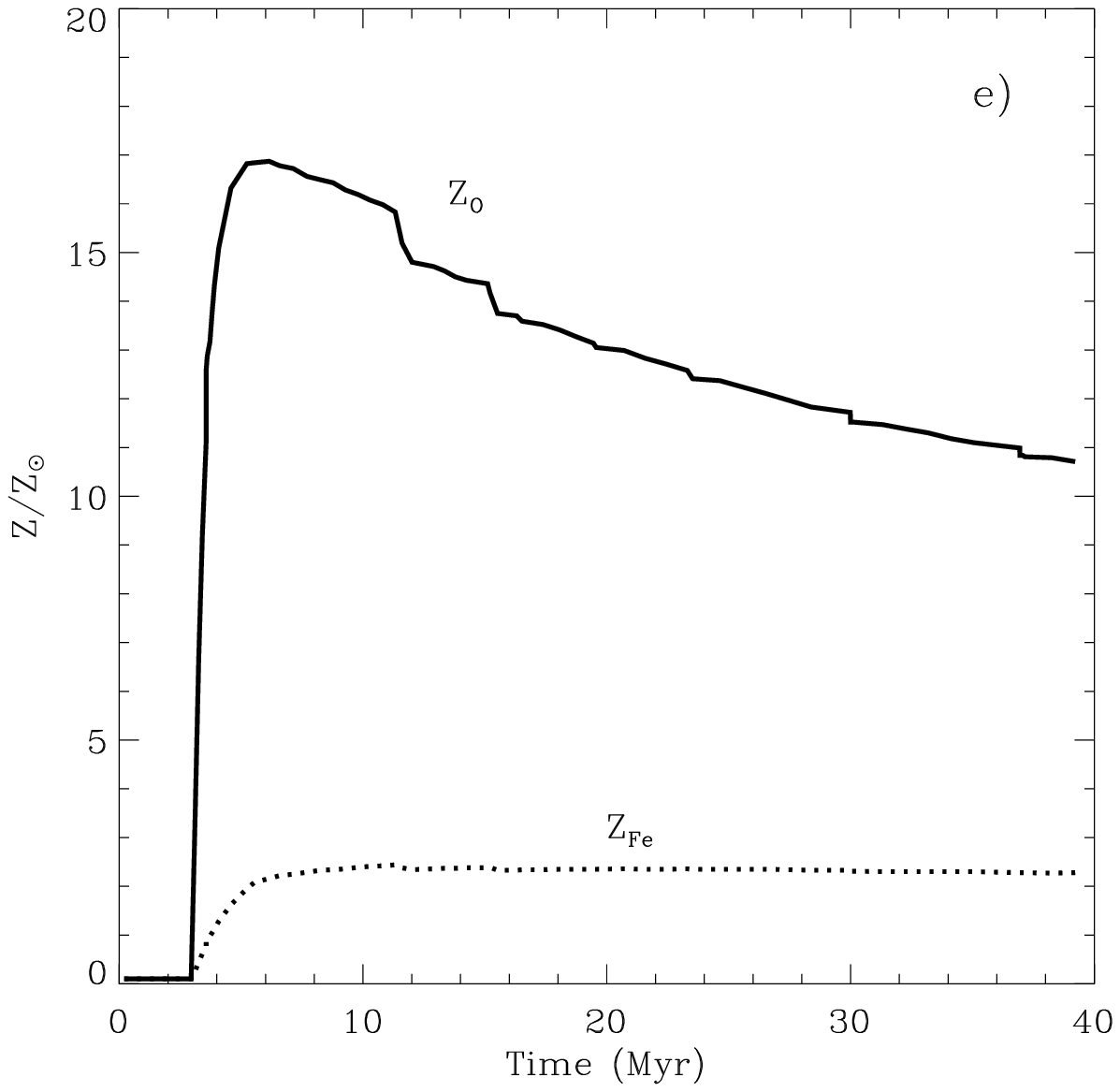,height=7cm,width=8cm,angle=0}
\hspace{2mm}
\psfig{figure=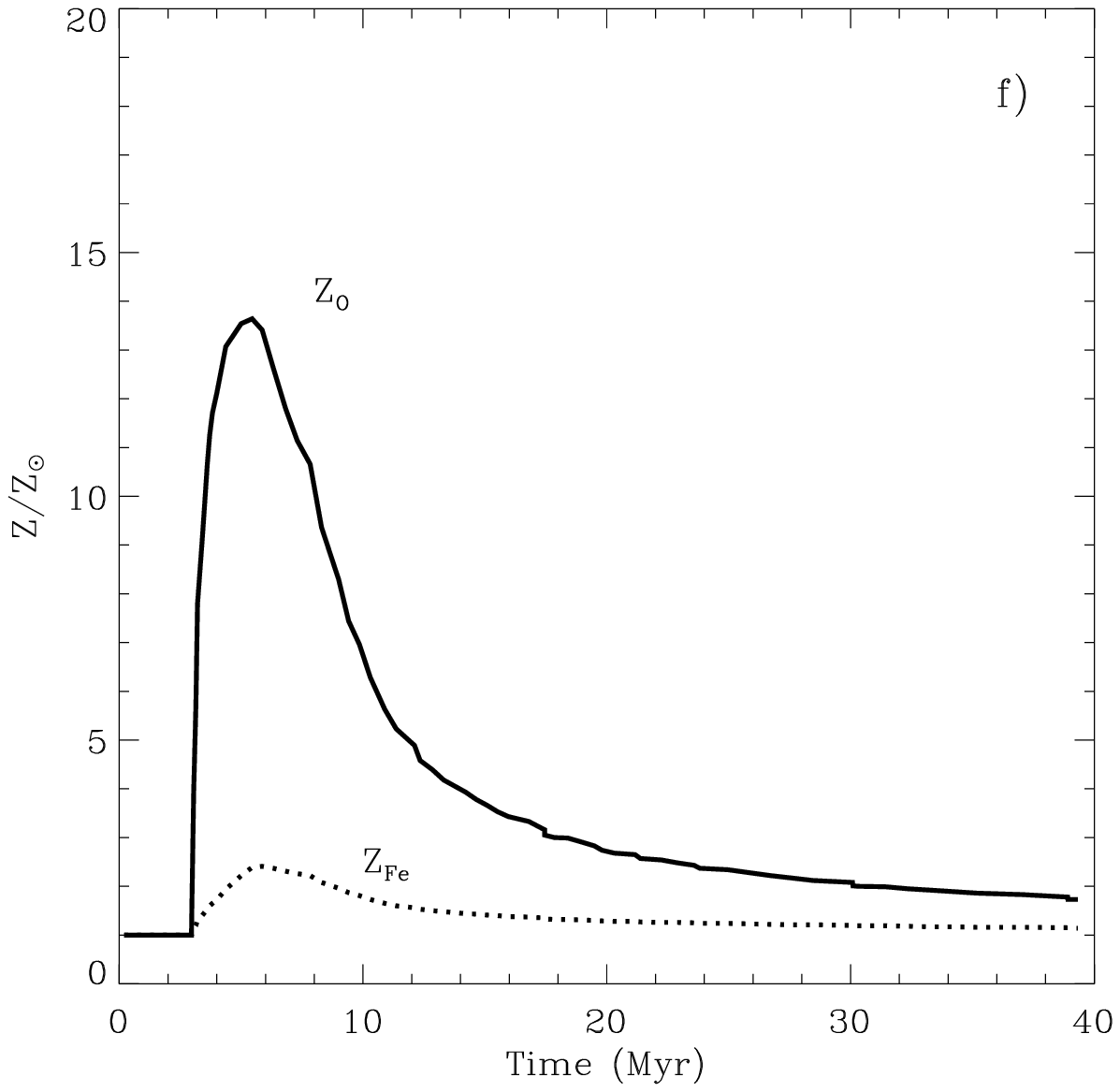,height=7cm,width=8cm,angle=0}}
}
\caption
{The panels display the oxygen and iron content of superbubbles (in solar 
units: Z$_O$ = 0.0083 and Z$_{Fe}$ =0.00126, from Grevesse and Noels 1996) for 
the models {\bf (a)}: 
1A (with thermal evaporation of the cold outer shell segments) and NW and 
A approximations; {\bf (b)} 1B with WW and T approximations; 
{\bf (c)} 2A (no mass evaporation, NW and A approximations); 
{\bf (d)} 2B, (no mass evaporation, WW and T approximations);
{\bf (e)} 3A, a 10$^7$M\sol starburst within a 0.1Z$_{\odot}$ ISM, and 
{\bf (f)} 3B, a 10$^7$M\sol starburst within a Z$_{\odot}$ ISM. 
}
\label{fig5} 
\end{figure*}

Figure~\ref{fig5} shows the evolution of metallicity of the superbubble 
interior as a 
function of time for the different model assumptions  regarding the yields, 
the mixing of heavy elements and the mass evaporation. 
Case 1A (using the NW and A approximations) rapidly (between 4 - 5 Myr) 
reaches over solar metallicities Z$_O$ $\sim$ 8 Z\sol for oxygen, and solar 
values for the iron tracer (Z$_{Fe}$) (Figure~\ref{fig5}a). Z$_O$ then slowly 
decays to about 3 Z\sol after 10~Myr of evolution, to solar after 20~Myr and 
to under solar ($\sim$ 0.5 Z\sol) at 
the end of the calculation. Z$_{Fe}$ becomes under solar after 6~Myr and
slowly approaches the metallicity of the host galaxy (0.1  Z\sol). The 
metallicity that results from  case 1B, assuming the WW and T approximations 
for the production of metals (Figure~\ref{fig5}b), reaches  
Z$_O$ $\sim$ 2 Z\sol after 
6 Myr of evolution, falling below solar metallicity just after 10 Myr to 
slowly approach 0.1 Z\sol at the end of the calculation. Z$_{Fe}$ on the other
hand, hardly shows a significant variation in this case and remains under 
solar throughout the calculation.

Case 2 (without mass evaporation) reaches, within short times  after the 
start 
of the SN phase (3 Myr), a value of Z$_{O}$ larger than 30 Z\sol (case 2A).
Even after a substantial fall, as 
the less massive stars become SN, this case ends with Z$_O \sim$ 15 Z\sol 
(Figure~\ref{fig5}c). 
In this case  Z$_{Fe} \sim$ 6 Z\sol after 
6~Myr of evolution and remains almost constant throughout the calculation. 
Case 2B (under the WW and T assumptions for the metal production rate)  
presents a rapid rise in Z$_O$ reaching a maximum Z$_O$ = 8 Z\sol at about 
10 Myr and then slowly declines to 6 Z\sol after 40 Myr. The Z$_{Fe}$ on the 
other hand, remains subsolar for the first 12 Myr and
then rapidly rises to reach 6 Z\sol after 40 Myr (Figure~\ref{fig5}d).

For the most energetic starburst (case 3) large values of Z$_O$ $\geq$ 10 
Z\sol are found within the first 15 Myr (Figure~\ref{fig5}e), when the NW and 
A metal production rates and Z$_{ISM}=0.1$Z$_{\odot}$ are used (model 3A). The 
enhanced metallicities however, are not diluted in this case  as the fast and 
hot quasi-adiabatic outer shell has inhibited mass evaporation. 

As we increase the initial ISM metallicity from 0.1\Zsol\ to 
\Zsol\ for model 3B, the interior bubble metallicity is lower and more peaked  
with time (Figure~\ref{fig5}f). This unexpected result is produced by a much 
faster shell cooling, as it is more metal rich, and thus the 
sarburst ejected metals are more efficiently diluted by the
evaporated gas from the shell.

Finally, we comment on present day abundance determination of X-ray gas.
 {\it ROSAT\/} observations allow, in principle, for the derivation of metal 
abundance, of a $(T\sim 1-30 \times 10^6\,\rm{K})$ plasma,  by
X-ray spectral fittings. However, the process is problematic, because of the 
poor
energy resolution of the {\it ROSAT-PSPC} detector, and the derived 
ISM metallicities
disagree, in many cases, with those deduced from optical observations
(Trinchieri \etal 1994; Saracco \& Ciliegi 1995; Bauer \&  Bregman 1996).
The situation has somewhat improved with the {\it ASCA\/} and {\it BeppoSAX\/} 
observations yet the X-ray measurements still suggest significantly sub-solar 
abundance for Fe combined with somewhat sub-solar abundances for 
$\alpha$-elements (Bauer \& Bregman 1996, Ptak \etal 1999, Persic \etal 1998). 
In some of the cases, the optical observations clearly indicate higher 
metallicities.

In other words, the complexity of the X-ray spectra of starburst 
galaxies, combined with the poor spectral resolution of the pre-Chandra 
and pre-XMM observations, makes the determination of metallicities  very model 
dependent (e.g. Netzer 1999). Dahlem, Weaver \& Heckman (1998) and Weaver 
\etal 
(1999) have discussed the difficulties in deriving metallicities from X-ray 
data, and, by combining  {\it ROSAT\/} and {\it ASCA\/} data for nearby edge-on
starbursts like NGC~253 and M~82, they concluded that the fitting is 
consistent with near solar abundances and therefore, extremely low 
metallicities as 
derived by only {\it ROSAT\/} or {\it ASCA\/} data are no longer required. 

\section{The X-ray luminosity of superbubbles}

The X-ray emission from superbubbles arise from two different physical 
regions: the dense outer shell of accelerated ISM and the hot superbubble 
interior  (Suchkov \etal 1994, Silich \& 
Tenorio-Tagle 1998, Strickland \& Stevens 1998, D'Ercole \& Brighenti 1999). 
Our  description of the inner bubble structure is based on the Weaver \etal 
(1977) similarity solution, and assumes that the density $n$ and temperature 
$T$ profiles can be approximated by
\begin{eqnarray}
      \label{eq.10}
      & & \hspace{-0.5cm}
n = n_c(1-x)^{-\lambda},
      \\[0.2cm]
      & & \hspace{-0.5cm}
T = T_c(1-x)^{\lambda},
\end{eqnarray}
where n$_c$ and T$_c$ are the central hot gas number density and temperature,
and $x=r/R_{sh}$ is the dimensionless distance from the bubble center. 
To account for the expected difference in density and temperature 
distributions for adiabatic and radiative shell segments, we
approximate the inner bubble luminosity as
\begin{equation}
      \label{eq.11}
L_{x,in} = \epsilon L_R + (1-\epsilon) L_A
\end{equation}
where $\epsilon$ is the ratio of the radiative surface segments to the
total remnant surface area, and L$_R$ and L$_A$ are the X-ray
emissions from the interior of the fully radiative and adiabatic
bubbles, respectively (Silich \& Tenorio-Tagle, 1998)
\begin{eqnarray}
      \label{eq.11}
L_{x,in} = &\int_0^{2\pi} {\rm d}\phi  \int_0^{\pi} \sin \theta 
           {\rm d}\theta  \int_0^{R_{cut}} \xi n^2(r) \Lambda_x(T)
           r^2 {\rm d}r \nonumber \\
= & 3 \xi \lambda^{-1} n_c^2 \Omega I(T_c, \lambda)
\end{eqnarray}
Here I(T$_c, \lambda$) is a dimensionless integral given by
\begin{eqnarray}
      \label{eq.12}
I(T_c, \lambda) = & \frac{1}{T_c} \int_{T_{cut}}^{T_c} \Lambda_x(T)
                  \left(\frac{T}{T_c}\right)^{(1-3\lambda)/\lambda}\nonumber \\
                &  \left[1 - \left(\frac{T}{T_c}\right)^{1/\lambda}
                  \right]^2  {\rm d}T,
\end{eqnarray}
R$_{cut}$ and T$_{cut}$ are the X-ray cut-off radius and temperature 
respectively.
We used self-similar power index $\lambda = 2/5$ for L$_R$, and 
$\lambda = 1/20$ for  a more homogeneous gas distribution within 
a bubble with an adiabatic hot shell.
The X-ray luminosity from the adiabatic shell segments is taken to be
\begin{equation}
      \label{eq.13}
L_{x,shell} = \xi \sum n_{shock}^2 \Lambda_x(T_{shock}) \Delta R
            {\rm d} \Sigma,
\end{equation}
where ${\rm d} \Sigma$ is the adiabatic segment surface area, and
$\Delta$R is the segment thickness.

\begin{figure}
\psfig{figure=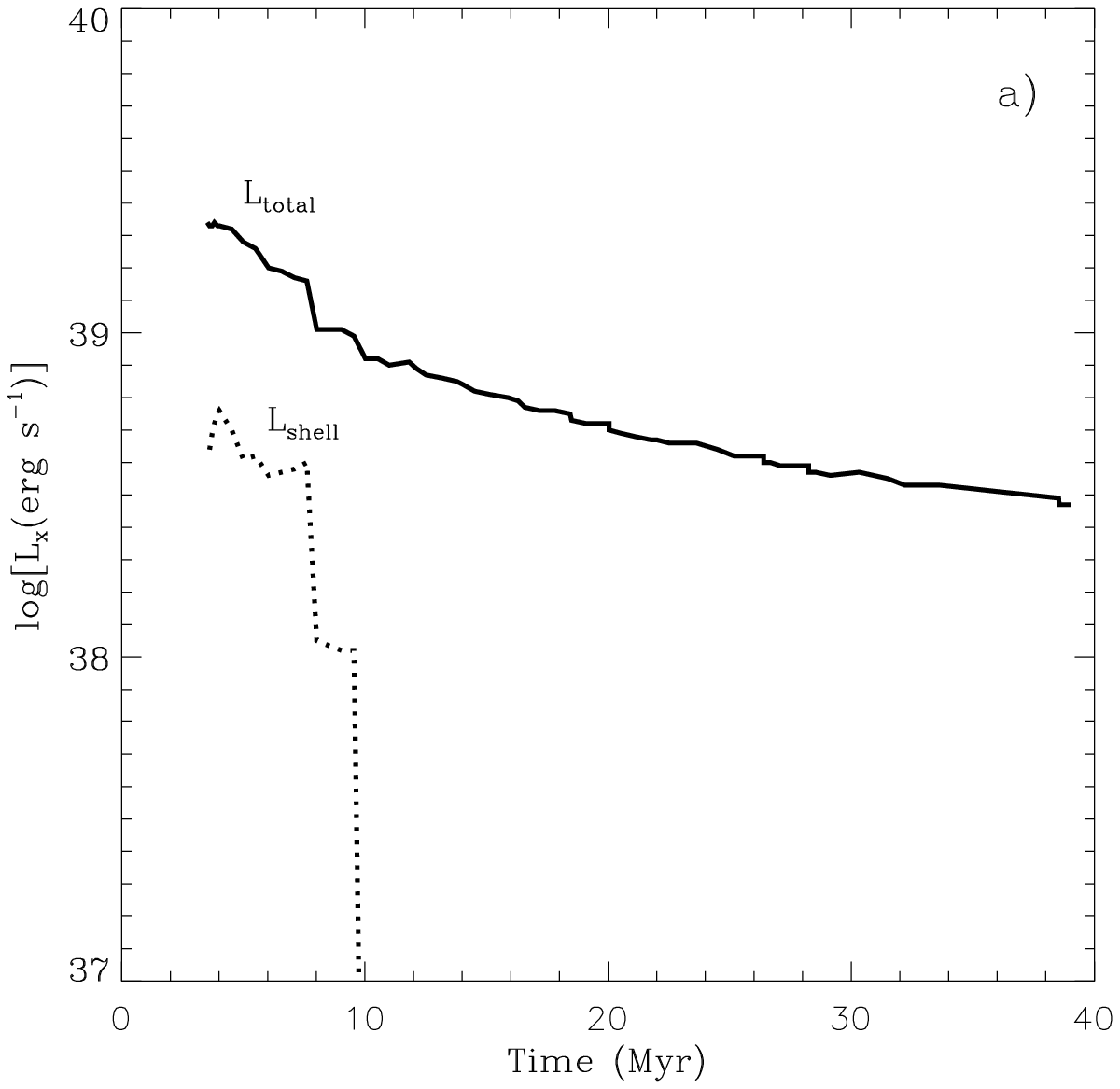,height=7cm,width=8cm,angle=0}
\psfig{figure=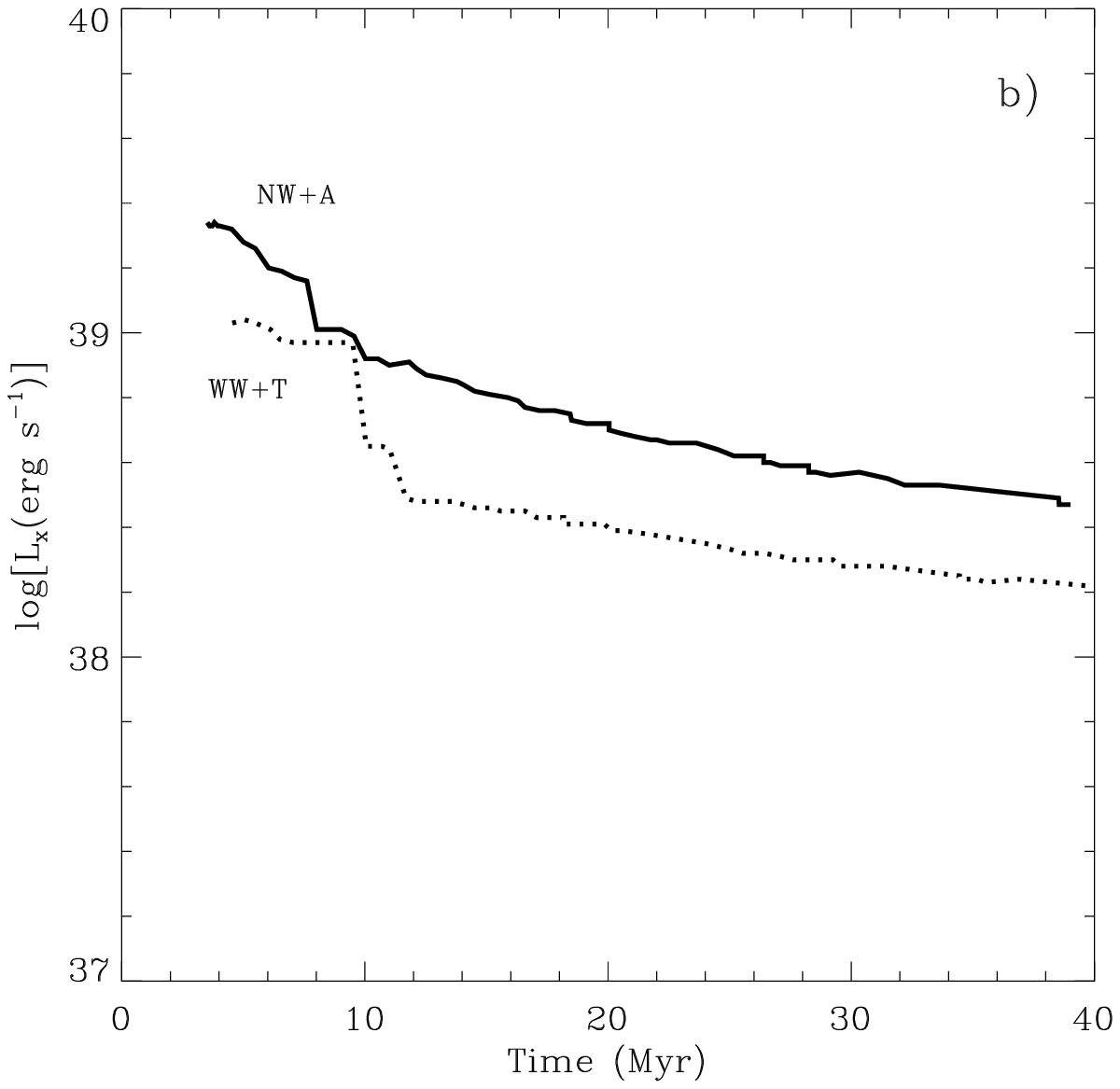,height=7cm,width=8cm,angle=0}
\psfig{figure=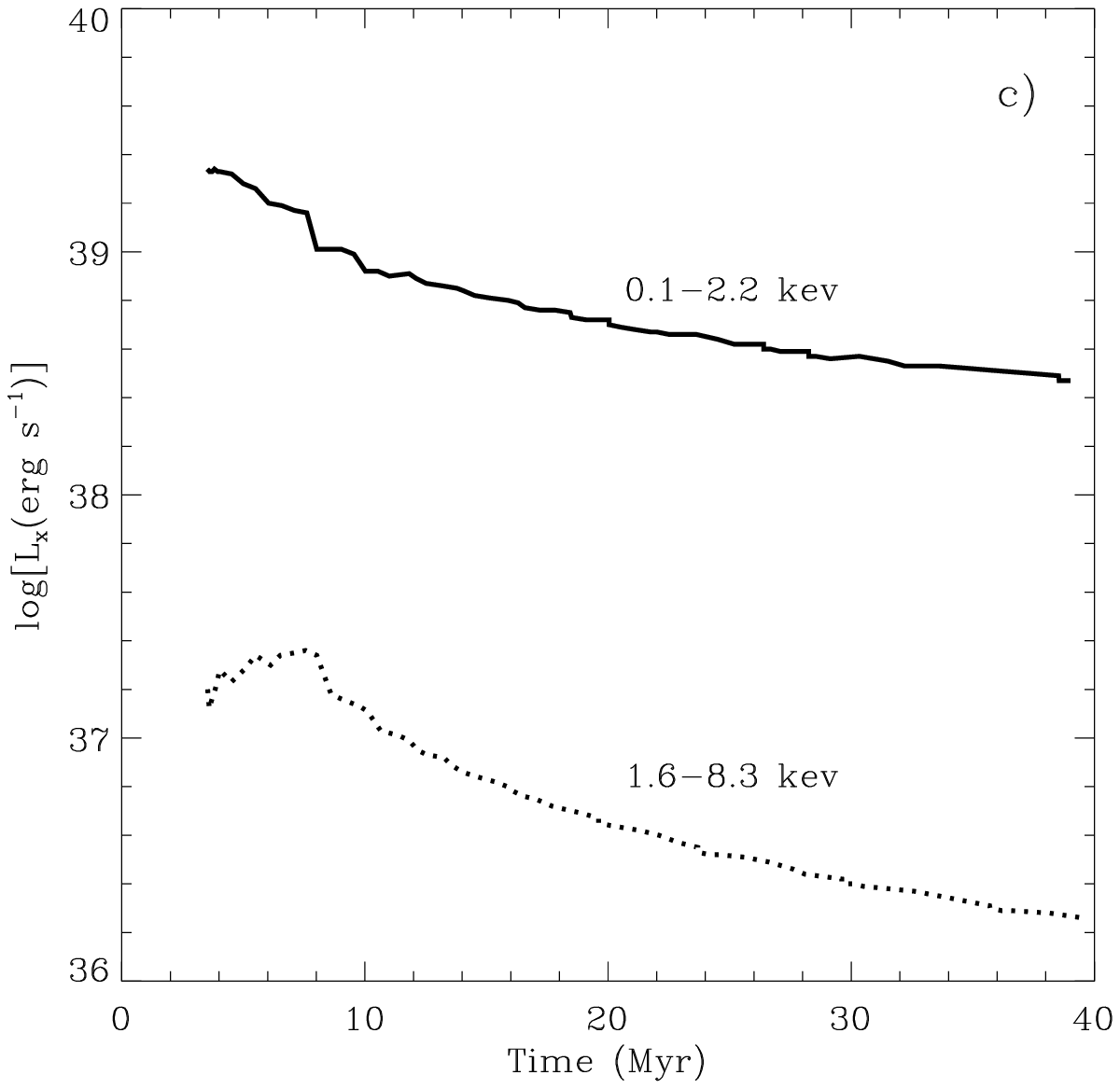,height=7cm,width=8cm,angle=0}
\caption
{The log  X-ray luminosity, as a function of time, 
for case 1A under  NW and  A approximations. {\bf (a)} shows the outer shell 
contribution (dotted line) and the total luminosity (solid line);
{\bf (b)} compares the above total luminosity with that derived for case 1B 
under the WW and T assumptions (dotted  line); and {\bf (c)} 
shows the X-ray luminosity in the two indicated X-ray bands for case 1A.
}
\label{fig6} 
\end{figure}

Figure~\ref{fig6}a shows the total and the shell contribution to the X-ray 
emissivity 
of case 1A, using the hot gas metallicities shown in Figure~\ref{fig5}a. The 
contribution of the shell to the total emission becomes most important
immediately after blow-out of the remnant from the 
central gas distribution into the 
extended halo of the host galaxy. During blow-out the shock reaches speeds of 
several hundreds of \kms\ leading to the high temperatures (T $\sim$ 1.4 
$\times$ 10$^7$ $v_{shock}^2$; where $v_{shock}$ is the shock velocity in 
units of 10$^3$ \kms) which allow the shocked gas to radiate in the X-ray 
regime. The shell contribution becomes rapidly less important as the shock 
slows down and the shell cools below the X-ray cut-off temperature ($\sim$ 5 
$\times$ 10$^5$) and condenses into a narrow and dense outer boundary of the 
superbubble. On the other hand, the luminosity of the superbubble interior 
is found to remain within the range of observed values (10$^{38}$ -- 5 
$\times$ 10$^{39}$ erg s$^{-1}$) throughout the evolution. Figure~\ref{fig6}b 
compares the time dependent X-ray luminosity for our case 1 under 
the two assumptions for metal production rate. Clearly, the lower 
metallicities reached in the WW model, lead to a smaller X-ray emission, 
although the difference never amounts to more than a factor of 5.

 Figure~\ref{fig6}c gives the contribution to the 
X-ray 
luminosity of case 1A in two different energy bands, as a function of time. 
The two energy bands considered are from 0.1 to 2.2, 
and from 1.6 to 8.3 keV (X-ray emissivities from Suchkov \etal 1994). 
Clearly, most of the X-ray luminosity arises from the  soft X-ray component. 
The  high energy  X-rays  are due mostly to  the larger 
central temperatures in the superbubble interior.

\begin{figure}
\psfig{figure=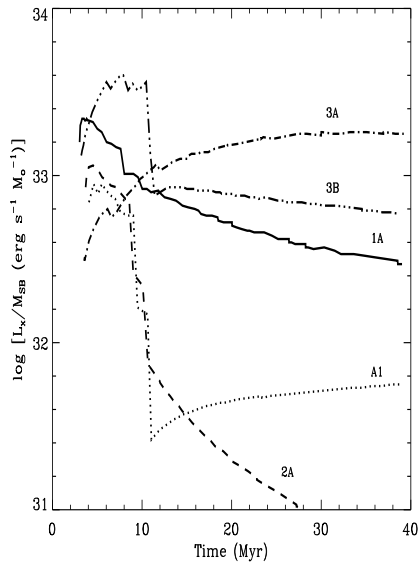,height=8cm,width=8cm,angle=0}
\caption
{The log of the X-ray luminosity of superbubbles per unit mass  as a 
function of time for cases 1A (low Z), comparison A1 (from paper I) (low Z), 
2A  (low Z), and massive 
starburst 3A (low Z) and 3B (solar abundance), as labelled. 
}
\label{fig7} 
\end{figure}

Figure~\ref{fig7} shows the time evolution of $L_{X}$, per unit 
mass,   in models 
1A, 2A, 3A and 3B, as well as the same quantity for model A1 from paper I which did 
not include  
changes in metallicity due to the injection of new metals into the
cavity. The main difference between
case 1A and  reference case A1 comes from the hot superbubble interior and can 
amount to more than an order of magnitude, particularly after 10~Myr, once the 
shell contribution has become negligible in both cases. Note that the X-ray 
luminosity of case A1 is in fact rather similar to that of case 2A (without 
thermal evaporation). However, in case 2 the reduced X-ray emissivity from 
the interior is due
to the fact that, without the substantial extra mass input, the 
amount of matter able to radiate is orders of magnitude too small. 
Case 3, our  most powerful starburst, leads to the largest 
luminosities of all cases. 

It is interesting to notice an important difference between the bubble
evolution in low (case 3A) and high (case 3B) metallicity ISMs.
Up to 10~Myr, the X-ray luminosity of the bubble in the metal-rich galaxy is almost 
an order of magnitude larger than that for the low metallicity one. However, 
as the high metallicity bubble  reaches the radiative phase 
much earlier than  the low metallicity one, 
the metal-rich shell (3B) cools rapidly around 11~Myr,
whereas in the  metal-poor case (3A) it remains adiabatic and hot. This
results in a rapid decrease of  L$_x$ in the high metallicity model,
which, after 12~Myr, drops below the X-ray emission from the low metallicity
one, and decline more slowly afterwards.    
The plot indicates that similar X-ray luminosities per unit mass ($\sim$ 10$^{33}$ 
erg s$^{-1}$ M\sol $^{-1}$) arise from
cases that allow  mass evaporation from the outer shell, irrespective
of the mass of the starburst causing the superbubble.
This is very different from the calculations that do not account for the 
change in metallicity of the superbubble interior
(case A1, from paper I) and also from those that do not account for mass 
evaporation (case 2). 

\section{Conclusions}

The calculations presented here, with the exception of those for the 
more massive starburst (case 3), differ only slightly in their final outcome 
regarding the size,  expansion velocity and amount of matter swept by  
 evolving superbubbles. However, they  show enormous differences in the 
metallicity of the hot
interior, and thus produce largely different X-ray emissivities.        

The effects of injection of new metals into the superbubble 
interior is most noticeable during the HII region lifetime (the first 10 Myr). 
If oxygen is used as a tracer, the metallicity reaches, 
immediately after the start of the SN phase, 
values well above solar metallicity.   
The maximum value is achieved  3 to 5 Myr after the beginning of this
phase. Later,
mixing with the evaporated mass  slowly reduces the impact of SNe.   
Note however, that if massive stars are assumed not to have strong winds, 
large values of $Z \geq$ \Zsol would remain present inside the superbubble  
for at least the time required to reach  the end of the 
type II SN phase (40 Myr).
This is true  even in the 
extreme case where the host galaxy 
initial metal abundance is well below solar.  
On the other hand, if iron is used as the tracer of metallicity, values  of 
Z$_{Fe}$ $\leq$ Z\sol are always predicted, regardless of the 
assumed iron production scheme.
The above results are to be compared to the cases without mass evaporation, 
all of  which led  to metallicities larger than solar throughout the evolution, 
regardless of the assumed tracer.

More massive starbursts result in very strong shocks and  longlasting 
hot, quasi-adiabatic outer shell from which thermal evaporation is strongly 
inhibited. Such starbursts show over solar metallicity throughout their 
evolution, regardless of the assumed tracer.      
     
The most important conclusion is that
enhanced metallicity of  superbubble interiors can strongly influence their
X-ray luminosity,
bringing theory and observations into a better agreement.

\section{Acknowledgments}
The authors acknowledge fruitful conversations with 
Andy Fabian, and with Dahlem and Weaver, who kindly provided us with their 
results prior to publication.

SAS acknowledges support from a Royal Society grant for joint projects
with countries of the former Soviet Union. SAS, GTT and ET also thank the IoA 
in Cambridge for their
support and hospitality during a visit to the United Kingdom where the
calculations were performed. HN acknowledges the hospitality of the IoA 
Cambridge,
where this collaboration was initiated, and financial support by the Israel 
Science Foundation. 
Research Grants from CONACYT (Mexican Research Council) are thankfully 
acknowledged by GTT and ET.


\begin{thebibliography}{} 


\bibitem{} Arnett, D. 1991, in Frontiers of Stellar Evolution, ed. D.L. Lambert
           (ASP Conf. Ser., 20), 389

\bibitem{} Bauer, F. \& Bregman, J.N. 1996, ApJ, 457, 382

\bibitem{} Bisnovatyi-Kogan, G.S. \& Silich, S.A. 1995, Rev. Mod. Phys.
           67, 661

\bibitem{} Brinks, E. \& Bajaja, E. 1986, A\&A 169, 14

\bibitem{} Chiosi, C., Nasi, E. \& Sreenivasan, S.P. 1978, A\&A,
           63, 103

\bibitem{} Chu, Y.-H., \& Mac Low, M.-M. 1990, ApJ, 365, 510

\bibitem{} Dahlem, M., Weaver, K.A. \& Heckman, T.M. 1998, ApJ Suppl. Ser.,
           118, 401

\bibitem{} D'Ercole, A. \& Brighenti, F. 1999, Astro-ph/9907005 

\bibitem{} Franco, J., Ferrara, A., Rozyczka, M., Tenorio-Tagle, G. \& 
           Cox, D.P. 1993, ApJ, 407, 100

\bibitem{} Heckman, T.M, Armus, L. \& Miley, G.K. 1990, ApJ Suppl. Ser.
           74, 833

\bibitem{} Heckman, T.M, Dahlem, M., Eales, S.A., Fabbiano, G., Weaver, K.
           1996, ApJ, 457, 616

\bibitem{} Heckman, T.M., Dahlem, M., Lehnert, M.D., Fabbiano, G.,
           Gilmore, D. \& Waller, W.H. 1995, ApJ, 448, 98 

\bibitem{} Heiles, C. 1979, ApJ, 229, 533

\bibitem{} Grevesse, N., Noels, A. \& Sauval, A.J. 1996, 
           ASP Conf. Ser., eds. Holt, S.S. \& Sonnebom, G. 99, 117

\bibitem{} Li, F. \& Ikeuchi, S. 1992, ApJ 390, 405

\bibitem{} Leitherer, C. \& Heckman, T.M. 1995, ApJS, 96, 9

\bibitem{} Martin, C.L. \& Kennicutt, R.C. 1995, ApJ, 447, 171

\bibitem{} Martin, C.L. 1996, ApJ, 465, 680

\bibitem{} Maeder, A. 1992, A\&A, 264, 105

\bibitem{} Maschenko, S., Thilker, D. \& Broun, R. 1999, A\&A, 343, 352 

\bibitem{} Meaburn, J. 1980, MNRAS, 192, 365

\bibitem{} Netzer, H., 1999, Astron.Nach., 4/5, 171

\bibitem{} Oey, M.S. 1996, ApJ, 467, 666

\bibitem{} Persic, M., Mariani, S., Cappi, M., Bassani, L., Danese, L., Dean, 
A.J., Di Cocco, G., Franceschini, A., Hunt, L.K., Matteucci, F., Palazzi, E.,
Palumbo, G.G.C., Rephaeli, Y., Salucci, P. \& Spizzichino, A., 1998, A\&A, 339,
L33

\bibitem{} Pilyugin, L.S. 1993, A\&A, 277, 42

\bibitem{} Pilyugin, L.S. \& Edmunds, M.G. 1996, A\&A, 313, 792

\bibitem{} Ptak, A., Serlemitsos, P., Yaqoob, T. \& Mushotzky, R. 1999, ApJS, 
120, 179
 
\bibitem{} Renzini, A., Ciotti, L., D'Ercole, A. \& Pellegrini, S.
           1993, ApJ, 419, 52

\bibitem{} Silich, S. 1995, Astron. and Astroph. Transactions,
           9, 85

\bibitem{} Silich, S. \& Tenorio-Tagle, G. 1998, MNRAS, 299, 249

\bibitem{} Stothers, R. 1972, ApJ, 175, 431

\bibitem{} Strickland, D.K. \& Stevens, I.R. 1998, MNRAS, 297, 747

\bibitem{} Suchkov, A., Balsara, D. Heckman, T. \& Leitherer, C. 1994, 
           ApJ 430,  511

\bibitem{} Tenorio-Tagle, G. 1996, AJ, 111, 1641 

\bibitem{} Tenorio-Tagle, G \& Bodenheimer, P. 1988, ARA\&A 26,145

\bibitem{} Tenorio-Tagle, G. \& Mu\~noz-Tu\~n\'on, C. 1998, 293, 299 

\bibitem{} Tenorio-Tagle, G, Silich, S., Kunth, D., Terlevich, E. \& 
Terlevich, R.: 1999, MNRAS, 309, 332

\bibitem{} Thielemann, F.-K., Nomoto, K., Shigeyama, T., Tsujimoto, T. 
           \& Hashimoto, M. 1992, in  Elements and the Cosmos, Eds. 
R. J. Terlevich, B. Pagel, R. Carswell \& M. Edmunds
Cambridge University Press, 68

\bibitem{} Thielemann, F.-K., Nomoto, K.
           \& Hashimoto, M. 1993, in Origin and Evolution of the Elements, 
Eds. N. Prantzon, E. Vangioni-Flan, M. Casse
Cambridge University Press, 297

\bibitem{} Trinchieri, G., Kim, D.-W.,
 Fabbiano, G. \& Canizares, C. R. C.  1994, ApJ 428, 555

\bibitem{} Tomisaka, K. \& Bregman, J.N. 1993, PASJ, 45, 513

\bibitem{} Walter, F., Kerp, J., Duric, N., Brinks, E. \&
           Klein, U. 1998, ApJ Letters, 502, L143  
\bibitem{} Wang, Q.D. \& Helfand, D. 1991, ApJ, 379, 327

\bibitem{} Wang, Q.D. 1999, ApJ Letters, 510, L39

\bibitem{} Weaver, K.A., Heckman, T. \& Dahlem, M., 1999,  astro-ph/9911446 

\bibitem{} Woosley, S.E., Langer, N. \& Weaver, T.A. 1993, ApJ, 411, 823


\end{thebibliography}
\end{document}